# IMAGE COMPRESSION USING ANTI-FORENSICS METHOD

M.S.Sreelakshmi[1] and D. Venkataraman[1]

[1]Department of Computer Science and Engineering, Amrita Vishwa Vidyapeetham, Coimbatore, India

mssreelakshmi@yahoo.com

d_venkat@cb.amrita.edu

## *ABSTRACT*

*A large number of image forensics methods are available which are capable of identifying image tampering. But these techniques are not capable of addressing the anti-forensics method which is able to hide the trace of image tampering. In this paper anti-forensics method for digital image compression has been proposed. This anti-forensics method is capable of removing the traces of image compression. Additionally, technique is also able to remove the traces of blocking artifact that are left by image compression algorithms that divide an image into segments during compression process. This method is targeted to remove the compression fingerprints of JPEG compression.*

## 1. INTRODUCTION

Due to wide availability of powerful graphic editing software image forgeries can be done very easily. As a result a number of forensics methods have been developed for detecting digital image forgeries even when they are visually convincing. In order to improve the existing forensics method anti-forensics methods have to be developed and studies, so that the loopholes in the existing forensics methods can be identified and hence forensics methods can be improved. By doing so, researchers can be made aware of which forensics techniques are capable of being deceived, thus preventing altered images from being represented as authentic and allowing forensics examiners to establish a degree of confidence in their findings.

## 2. OVERVIEW OF PROPOSED METHOD

When a digital image undergoes JPEG compression it's first divided into series of 8x8 pixel blocks. Discrete Cosine Transform (DCT) of each 8x8 is then calculated. From left to right, top to bottom, the DCT is applied to each block. DCT of an image is calculated using the equation

$$D(i,j)=\frac{1}{\sqrt{2N}}C(i)c(j)\sum_{x=0}^{N-1}\sum_{y=0}^{N-1}P(x,y)\cos[\frac{(2x+1)i\Pi}{2N}]\cos[\frac{(2y+1)j\Pi}{2N}]$$

Each DCT coefficients are then compressed through quantization. DCT coefficients are quantized by dividing each DCT coefficients by its corresponding quantization matrix (Q). Let X be a DCT coefficient at position (i,j) is quantized to X' such that X'=Round(X/Qi,j).Finally the quantized DCT coefficients are rearranged in zigzag order.

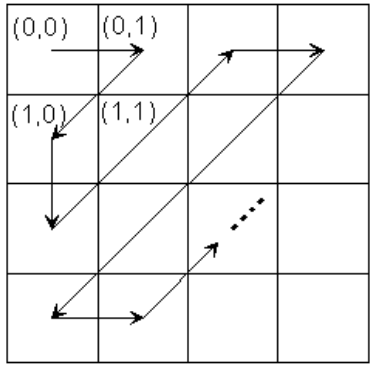

To decompress the image the sequence of quantized DCT coefficients are rearranged into its original order. Then dequantization is performed by multiplying each quantized DCT coefficients by its corresponding quantization matrix, Y = Qi,j * X'. Inverse DCT (IDCT) is applied to each of the DCT blocks and the resulting pixel value is rounded to nearest integer.

Since JPEG is lossy compression two types of artifacts are introduced into the compressed image transform coefficient quantization artifact and blocking artifact. Due to the quantization and dequantization operation the values of each DCT coefficients will be multiple of quantization stepsize Qi,j. This type of artifact is known as transform coefficient quantization artifact. The second compression artifact is blocking artifact which are introduced due pixel value discontinuity occurring across the block boundaries as a result of the block wise lossy compression [1].

## 2. FORENSICS METHODS TO DETECT JPEG COMPRESSION

JPEG image format is popularly used in most digital cameras and image processing software. Because of the lossy nature of JPEG compression, two important artifacts will be introduced into the image. The coupling of the quantization and dequantization operations force the value each DCT coefficient to be an integer multiple of the quantization step size. A second compression artifact is the pixel value discontinuities which occur across block boundaries as a result of the blockwise lossy compression employed by JPEG. These inconsistencies if detected could be used to check image integrity. The blocking artifact inconsistency found in a given image may tell the history that the image has been undergone [6].

### 2.1 BLOCKING ARTIFACT MEASURE

Blocking artifact value is introduced in JPEG compressed images. Thus the blocking artifact value can be used to detect image forgeries [6].

Blocking artifact for each block is estimated via:

$$B(i) = \sum_{k=1}^{64} | D(k) - Q(k) round(\frac{D(k)}{Q(k)}) |$$

Where, B (i) is the estimated blocking artifact for the testing block i, and D(k) is the DCT coefficient at position k. Q(1:64) is the estimated DCT quantization table.

The procedure of the quantization table estimation is:

(1) Calculating DCT coefficients of each 8X8 image block.

(2) Calculating the power spectrum (P) of the histogram of DCT coefficients for each of the 64 frequencies.

(3) Calculating the second derivative of P, and then low-pass filtering it.

(4) Calculating the local minimum number (Num) of the filtered second derivative of P.

(5) The estimated quantization step of the DCT frequency is estimated as Num +1.

The blocking artifact measure (BAM) for the whole image is then calculated based on the blocking artifacts of all blocks:

$$BAM = \frac{1}{n} \sum_{i} B(i)$$

Where n is the total number of image blocks. Blocking artifact value for an uncompressed image will be zero but for compressed image it will be nonzero value.

## 4. LITERATURE SURVEY

### 4.1 FORENSICS METHODS

In paper [7], proposed method derives a new, maximum likelihood estimate of the Laplacian parameter using the quantized coefficients available at the decoder. The benefits of biased reconstruction can be quantified through extensive simulations. It's demonstrated that such improvements are very close to the best possible resulting from centroid reconstruction. Assuming a Laplacian distribution for the unquantized, AC DCT coefficients, derive the ML estimate of the Laplacian parameter using only the quantized coefficients available to the decoder. This estimate gives modest improvements in PSNR.

In paper [6], propose a passive way to detect digital image forgery by measuring its quality inconsistency based on JPEG blocking artifacts. A new quantization table estimation based on power spectrum of the histogram of the DCT coefficients is firstly introduced, and blocking artifact measure is calculated based on the estimated table. The inconsistencies of the JPEG blocking artifacts are then checked as a trace of image forgery. This approach is able to detect spliced image forgeries using different quantization table, or forgeries which would result in the blocking artifact inconsistencies in the whole images, such as block mismatching and object retouching.

In paper [4], a method was developed for the reliable estimation of the JPEG compression history of a bitmapped image. Not only an efficient method was presented to detect previous JPEG compression but also a very reliable MLE method was devised to estimate the quantizer table used. The detection method can trace JPEG images which are visually undistinguishable from the original and is extremely reliable for higher compression ratios, which is the range of interest. Detection can be made with QF as high as 95. It is likely that there will be no need for further processing the image for high QF, so that it is more important to accurately identify the high-compression cases. Our method has not failed yet in those circumstances.

### 4.2 ANTI FORENSICS METHODS

It's possible that image manipulators can be done undetectably using anti-forensics counter measure. In paper [5], it's possible two represent a previously JPEG compressed image as never compressed, hide evidence of double JPEG compression, and falsify image's origin. Simple anti-forensics methods have been developed to render JPEG blocking artifact both visually and statistically undetectable without resulting in forensically detectable changes to an image. This technique can be used to fool forensic algorithm designed to detect evidence of prior application of JPEG compression within uncompressed image, determine an image/s origin, detect multiple application of JPEG compression, and identify cut and paste type image forgeries

In paper [2], propose an anti-forensics operation capable of disguising key evidence of JPEG compression. It operates by removing the discrete cosine transform (DCT) coefficient quantization artifacts indicative of JPEG compression. The resulting anti-forensically modified image can then be re-compressed using a different quantization table to hide evidence of tampering or to falsify the images origin. Alternatively, further processing can be performed to remove blocking artifacts and the image can be passed off as never-compressed. This is accomplished by adding noise to the set of quantized DCT coefficients from a JPEG compressed image so that the distribution of anti-forensically modified coefficients matches an estimate of the distribution of unquantized DCT coefficients.

In paper [3], propose anti-forensics methods to removing the artifacts which wavelet-based compression schemes introduce into an image's wavelet coefficient histograms. After anti-forensics operation is applied, an image can be passed off as never compressed, thereby allowing forensic investigators to be misled about an image's origin and processing history. This technique operates by adding anti-forensics dither to the wavelet coefficients of a compressed image so that the distribution of anti-forensically modified coefficients matches a model of the coefficients before compression.

## 5. IMPLEMENTATION DETAILS

When a digital image undergoes JPEG compression it's first divided into series of 8x8 pixel blocks. Discrete Cosine Transform (DCT) of each 8x8 is then calculated. Each DCT coefficients are then compressed through quantization. DCT coefficients are quantized by dividing each DCT coefficients by its corresponding quantization matrix (Q).Finally the quantized DCT coefficients are rearranged in zigzag order. To decompress the image the sequence of quantized DCT coefficients are rearranged into its original order. Then dequantization is performed by multiplying each quantized DCT coefficients by its corresponding quantization matrix. Inverse DCT (IDCT) is applied to each of the DCT blocks and the resulting pixel value is rounded to nearest integer.

Since JPEG is lossy compression two types of artifacts are introduced into the compressed image transform coefficient quantization artifact and blocking artifact. Due to the quantization and dequantization operation the values of each DCT coefficients will be multiple of quantization step size $Q_{i,j}$. This type of artifact is known as transform coefficient quantization artifact. The second compression artifact is blocking artifact which are introduced due pixel value discontinuity occurring across the block boundaries as a result of the block wise lossy compression [1].

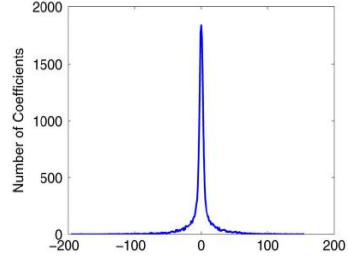

Figure 5.1: Histogram of DCT coefficients from an uncompressed image

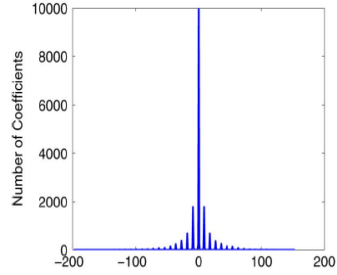

Figure 3.2: Histogram of DCT coefficients from the same image after JPEG compression

## 5.1 COMPRESSION ARTIFACT REMOVAL

Anti-forensics compression starts with the same steps of JPEG compression. Take the DCT coefficients of the image and add anti forensics dither (N) to each of the DCT coefficients. Each anti-forensically modified DCT coefficient Z is obtained according to the equation [1].

$$Z = Y + N$$

Where, N is anti-forensics dither. Choice of the value of N is critical. If the value of noise added is less than quantization artifact will be remained. Where as if the value of additive noise is too high visual distortion may be introduced. The value of N is determined by the value of the DCT coefficients [1]. If the DCT coefficients are zero then value of N is determined by

$$p(N=n|Y=0) = \begin{cases} \frac{1}{c_0} e^{-\lambda_{ML}|n|} & \text{if } \frac{-Q_{i,j}}{2} > n > \frac{Q_{i,j}}{2} \\ 0 & \text{otherwise} \end{cases}$$

If the DCT coefficients are nonzero then value of N is determined by

$$p(N=n|Y=y) = \begin{cases} \frac{1}{c_1} e^{-\sin(y)\lambda_{ML}(n+\frac{q}{2})} & \text{if } \frac{-Q_{i,j}}{2} > n > \frac{Q_{i,j}}{2} \\ 0 & \text{otherwise} \end{cases}$$

Where, N is the total number of observations of the quantized (i, j) DCT coefficient, $N_0$ is the number of observations taking the value zero, and $N_1$ is the number of nonzero observations.

$c_0 = 1 - e^{-\lambda_{ML} Q_{i,j}/2}$.

$$c1 = \frac{1}{\lambda_{ML}}(1 - e^{-\lambda_{ML}*Q_{i,j}})$$

$$\lambda_{ML} = -\frac{2}{Q_{i,j}}\ln(\gamma)$$

$$\gamma = \frac{-N0Q_{i,j}}{2NQ_{i,j}+4S} + \frac{\sqrt{N0^2 Q_{i,j} - (2N_1 Q_{i,j} - 4S)(2NQ_{i,j} + 4S)}}{2NQ+S}$$

$$S = \sum_{k=1}^{N} |y_k|$$

This choice of conditional noise distributions ensures that the distribution of anti-forensically modified DCT coefficients will exactly match the model distribution of unmodified DCT coefficients [1].

Image forgery like cut and paste forgery can be done using this anti-forensics method. Anti-forensics compression can be applied to cut and paste image so that it won't be detected as forged image [5].

## 6. EXPERIMENTAL RESULTS

Results obtained using anti-forensically compressed image are shown in Fig.6.1, which displays the JPEG compressed image before and after anti-forensics algorithm is applied. Both the histogram of DCT coefficients of uncompressed image and anti-forensically compressed image are same. Similarly blocking artifact value of anti-forensically compressed image is zero, shown in Fig 6.3. Image forgery such as cut and paste forgery can also be done using this method. Results of forged image are also given in Fig 6.2. PSNR value of anti-forensically compressed image and original image is 62.13 db which indicate that the proposed anti-forensics introduce acceptable level of distortion.

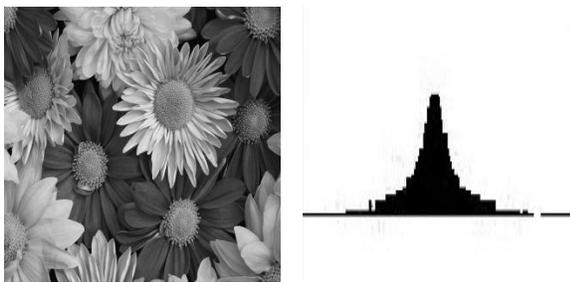

Fig 6.1(a) Uncompressed image and histogram of uncompressed image's DCT coefficients

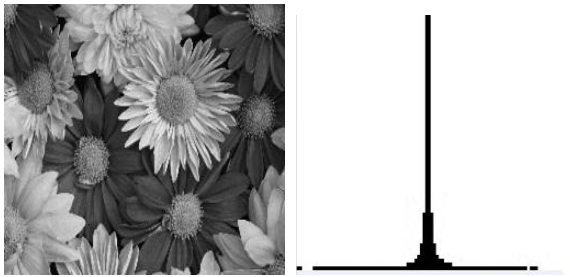

Fig 6.1(b) JPEG compressed image and its histogram of DCT coeffcients

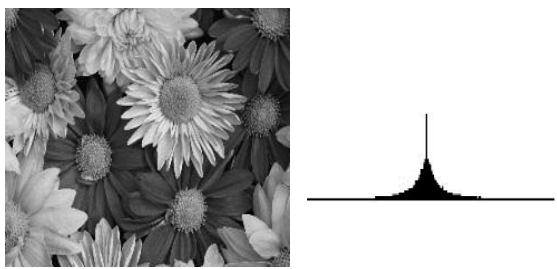

Fig 6.1(c)Anti-forensics compressed image and its histogram of DCT coefficients

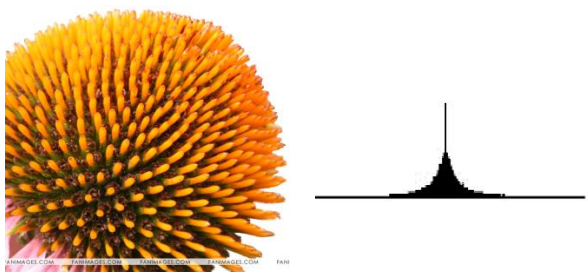

Fig 6.2(a) Original image and its histogram of DCT coefficients

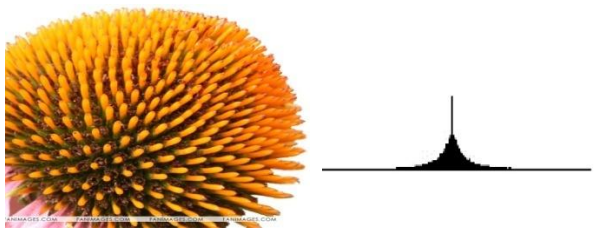

Fig 6.2 (b) Forged images (edited the text in rightmost end and compressed using anti-forensics method) and histogram of its DCT coefficients histogram

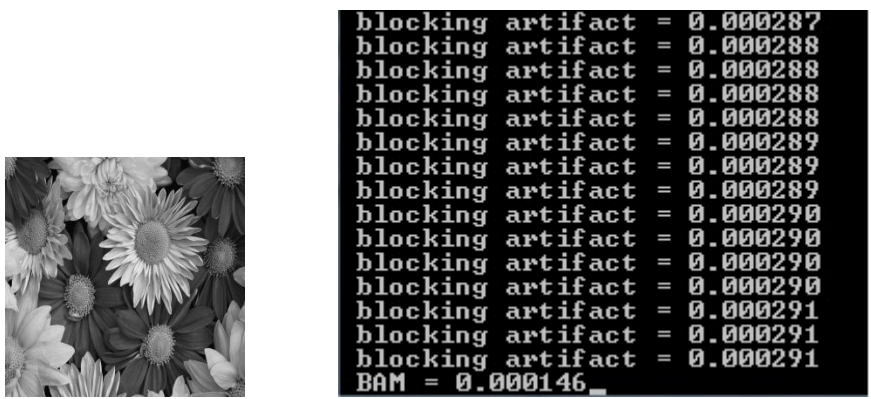

Fig 6.3 (a) Blocking artifact value of uncompressed image

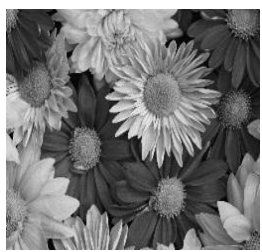

Fig 6.3 (b) Blocking artifact value of JPEG compressed image

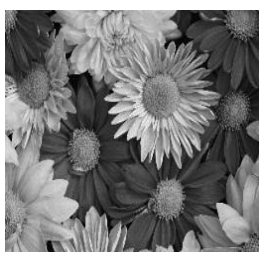

Fig 6.3 (c) Blocking artifact value of anti-forensics compressed image

## CONCLUSION

Image is compressed using anti forensics method. Two forensic methods for detecting image compression, histogram of the DCT coefficients and blocking artifact measure have been calculated to determine whether the image has been compressed or not. Both the histogram and blocking artifact value are same as the uncompressed image. Hence it's not detected as compressed image. These are the loopholes in the existing forensic method. Hence the forensic methods have to be improved.

**Authors : M.S.Sreelakshmi** was born in Kerala, India, in 1988. She received the B.Tech degree in computer science and engineering from the University of kerala, Kerala in 2010. She has been working towards the M.Tech degree in Computer Vision and Image Processing under the Department of Computer Science, Amrita Vishwa Vidyapeetham, Coimbatore, India.